# Energetic electron enhancements below the radiation belt and X-ray contamination at low-orbiting satellites

Alla V. Suvorova[1,2], Alexei V. Dmitriev[1,2], and Chien-Ming Huang[1]

[1] Institute of Space Science, National Central University, No 300 Jungda Rd, Jhongli, Taoyuan 32001,Taiwan
[2] Skobeltsyn Institute of Nuclear Physics, Lomonosov Moscow State University, Moscow 119234, Russia

Correspondence should be addressed to Suvorova Alla; alla@jupiter.ss.ncu.edu.tw

**Abstract.**

The work concerns a problem of electron-induced contaminant at relatively low latitudes to high-energy astrophysical measurements on board the low-orbiting satellites. We show the results of a statistical analysis of the energetic electron enhancements in energy range 30-300 keV observed by a fleet of NOAA/POES low-orbiting satellites over the time period from 1999 to 2012. We demonstrate geographical distributions of great and moderate long-lasting enhancements caused by different type of the solar wind drivers.

## 1. Introduction

Instrumental measurements in high-energy astrophysics require knowledge of contaminating background radiation of local (magnetospheric) origin [1]. Either by direct penetration or by secondary radiations produced in payload materials, photon detectors may at times give spurious responses, particularly if the "background" radiations are nonsteady [2]. It was recently pointed out that the most important effect limiting the accuracy of the cosmic X-ray background measurements is related to the intrinsic background variation in detectors [3,4]. This problem was comprehensively discussed at the Workshop on Electron Contamination in X-Ray Astronomy Experiments in 1974 [5]. It was shown that detectors of X- and gamma-ray on sounding rockets, on balloons and on board the low-orbiting satellites are subject to in-orbit enhanced background noise caused by the magnetospheric electrons, especially in the Earth's auroral zone and zone of trapped radiation (radiation belt), that is, at high latitudes, and also in South Atlantic Anomaly (SAA) at low latitudes (see also [6,7]). To minimize the contamination, most cosmic and galactic X- and gamma-ray measurements made from within the magnetosphere are conducted at equatorial or low-to-middle magnetic latitudes, where the influence of auroral and radiation belt effects is expected to be small. It is considered that data below the radiation belt outside the region of SAA are particularly valuable to satellite missions in high-energy astrophysics. However, even then during the early 1970s, the X-ray astronomers unexpectedly revealed the electron-induced contaminant at relatively low latitudes as well, which was a few times higher than the cosmic X-ray background [2, 8]. They found several events when flux intensity of electrons with energy of tens of keV was as large as $\sim 10^3$ el/(cm$^2$ s sr)$^{-1}$, exceeding quiet level by 2 orders of magnitude. Note that it is still much lower than in the radiation belt and auroral zones (see [9]). In addition to astrophysical measurements, ionospheric and atmospheric studies [10-17] and satellite data failures studies (e.g., [18,19]) also found several effects suggesting that electron impact is important factor at low and middle latitudes. That is, more importantly, the occasional electron flux increases below the radiation belt were discovered even earlier in direct satellite-borne measurements [20,21] and then corroborated in several studies [22-25]. They reported about sporadic fluxes of very high intensity which was comparable with the auroral precipitation. However, the direct observations of sporadic events caused strong argument due to a doubt about validity of measured high intensity (see review by Paulikas [26]). As a result of this, despite importance of low-latitude measurement of electron fluxes recognized earlier [5], further investigation of the enhanced electrons phenomenon was not carried out.

Until recently, sporadic enhancement of energetic electrons below the inner radiation belt (IRB) was a poor-studied phenomenon [15,27]. Comprehensive studies based on large statistics collected for more than ten years [28-31] have showed that fluxes of quasitrapped electrons within the energy range 10-300 keV can increase dramatically by a few orders of magnitude relative to the quiet level at very low L shells ($L < 1.1$), in a region called a forbidden zone. The most extreme intensity of forbidden zone fluxes of the order of auroral precipitation, $\sim 10^6$-$10^7$ (cm$^2$ s sr)$^{-1}$, was observed during some major storms driven by a coronal mass ejection (CME). Nevertheless, CME-type or major storms themselves are not a necessary condition for electron enhancements in the forbidden zone. Another important solar wind driver resulting in significant flux enhancements is the extremely strong solar wind dynamic pressure, as it occurred on 21 January 2005 [32]. It can be easily understood that large enhancements occur much less frequently than moderate ones. The moderate fluxes are smaller by one-two orders of magnitudes. They are mostly associated with low-to-moderate level of geomagnetic activity and minor storms (major storms can also contribute, though). Minor storms, as known, are mainly driven by corotating interaction regions (CIR) and high speed solar wind streams (HSS) [33].

This paper describes the results of a statistical analysis of the energetic electron enhancements observed by a fleet of NOAA/POES low-orbiting satellites over the time period from 1999 to 2012. We demonstrate geographical distributions of great and moderate enhancements caused by different type of the solar wind drivers.

## 2. Data from NOAA Satellites

We used time profiles of 30-300 keV electron fluxes measured on board the polar orbiting NOAA/POES satellites [34]. The POES satellites have Sun-synchronous orbits at altitudes of

~ 800 - 850 km (with ~100 minute periods of revolution). It is well known that the electron measurements can be distorted by proton contamination and nonideal detector efficiency. According to a comprehensive study [35] the 30 keV electron fluxes should be, on an average, more than two times larger than uncorrected fluxes. The 100 keV electron fluxes practically do not change, while the 300 keV electron fluxes should be decreased by about twenty percent. Because this factor is not crucial for the current study, we present uncorrected fluxes.

### 3. Enhancements of the Quasitrapped Energetic Electrons

Figure 1 presents geographical distributions of energetic electron fluxes in three energy ranges : >30 keV, >100 keV, and >300 keV. The data was compiled from measurements by two orthogonally oriented detectors (0°-detector and 90°-detector) of three POES satellites (NOAA-15, NOAA-16, NOAA-17) during the major (CME-driven) geomagnetic storm on 15-16 May 2005. In each spatial bin, the maximal value of flux instead of averaged one was used. The flux intensities below IRB in all three energy bands exceeded the quiet level by ~5-6 orders of magnitude. The lower energy electrons outside the SAA region achieved an extremely large value of $3 \cdot 10^7$ (cm$^2$ s sr)$^{-1}$, as much as in the IRB (including SAA) and auroral zone with outer radiation belt. The enhancements in >30 keV and >100 keV were long-lasting. The most prolonged equatorial enhancement in >30 keV occupied the forbidden zone at L shell of 1.05 – 1.15 for more than 20 hours.

Figure 2 presents geographical distributions of energetic electron fluxes during the prolonged compression of the magnetosphere by extremely high solar wind dynamic pressure of more than 150 nPa [32]. The magnetic storm was of moderate strength. Due to the compression, the Earth's magnetopause shrunk to about ~3-4 Re in the subsolar region, radiation belt and ring current moved closely to the Earth. The enhancement of >30 keV electrons was observed during 6 hours. Fluxes of electrons with higher energies were also increased, but due to fast azimuthally drift they disappeared in one or two hours.

Figure 3 demonstrates that >30 keV electrons can appear even during weak geomagnetic storms. The global map was compiled from measurements by five NOAA/POES satellites for one year 2008. During this year of solar activity minimum there were only minor (CIR/HSS-driven) storms of intensities less than 50 nT. In course of the year there were 60 days when the electron fluxes below IRB increased to ~$10^4$ (cm$^2$ s sr)$^{-1}$. However, even moderate electron enhancements can significantly contaminate to X-ray background.

## 4. Summary

In this paper we are concerned with a very important problem of the electron contamination to high-energy astrophysical measurements. The study is based on long-term statistics of the energetic electron observations by low-orbiting satellites. We have demonstrated three cases of electron fluxes that significantly exceeded a quite level: a major geomagnetic storm, a strong compression of the magnetosphere, and one-year period of the solar activity minimum leading to a weak geomagnetic activity.

The phenomenon of "forbidden zone electron" relates to the magnetospheric electric fields driven by external parameters, the solar wind, and interplanetary electric field [31]. A notable feature of the "forbidden-zone" 30 keV electrons is their long persistence for about several hours. It is important that the significant and longtime electron enhancements at equatorial latitudes occur quite often during moderate CIR/HSS-storms.

**Conflict of Interests**
The authors declare that there is no conflict of interests regarding the publication of this paper.

**Acknowledgements**
The authors thank a team of NOAA's Polar Orbiting Environmental Satellites for providing experimental data about energetic particles. The work of Alla V. Suvorova was supported by grant NSC-102-2811-M-008-045 from the National Science Council of Taiwan. Alla V. Suvorova and Alexei V. Dmitriev gratefully acknowledge the support of part of this work by Grant NSC103-2923-M-006-002-MY3/14-05-92002HHC_a from Taiwan-Russia Research Cooperation.

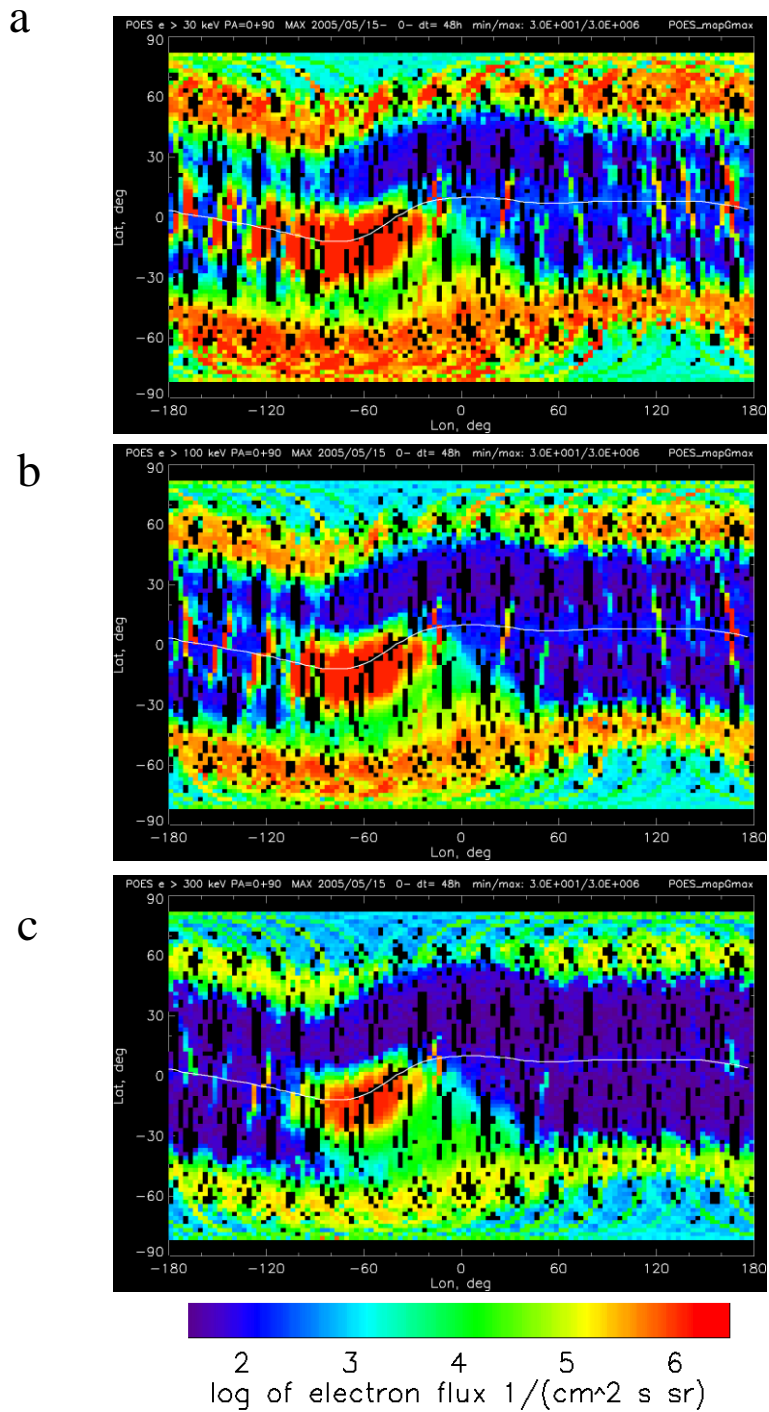

**Figure 1.** Energetic electron enhancement during the major geomagnetic storm on 15-16 May 2005. Global distribution of the electron fluxes in energy range: (a) >30 keV; (b) >100 keV; (c) >300 keV measured by the NOAA/POES -15, -16, -17 satellites at altitude of ~850 km. The maps are composed of data retrieved from two orthogonally directed detectors (see details in the text). The white curve indicates the geomagnetic equator. Intensity of energetic electron fluxes extremely and globally enhances at equator-to-low latitudes (IRB and below it) even exceeding one at high latitudes (ORB and auroral zone).

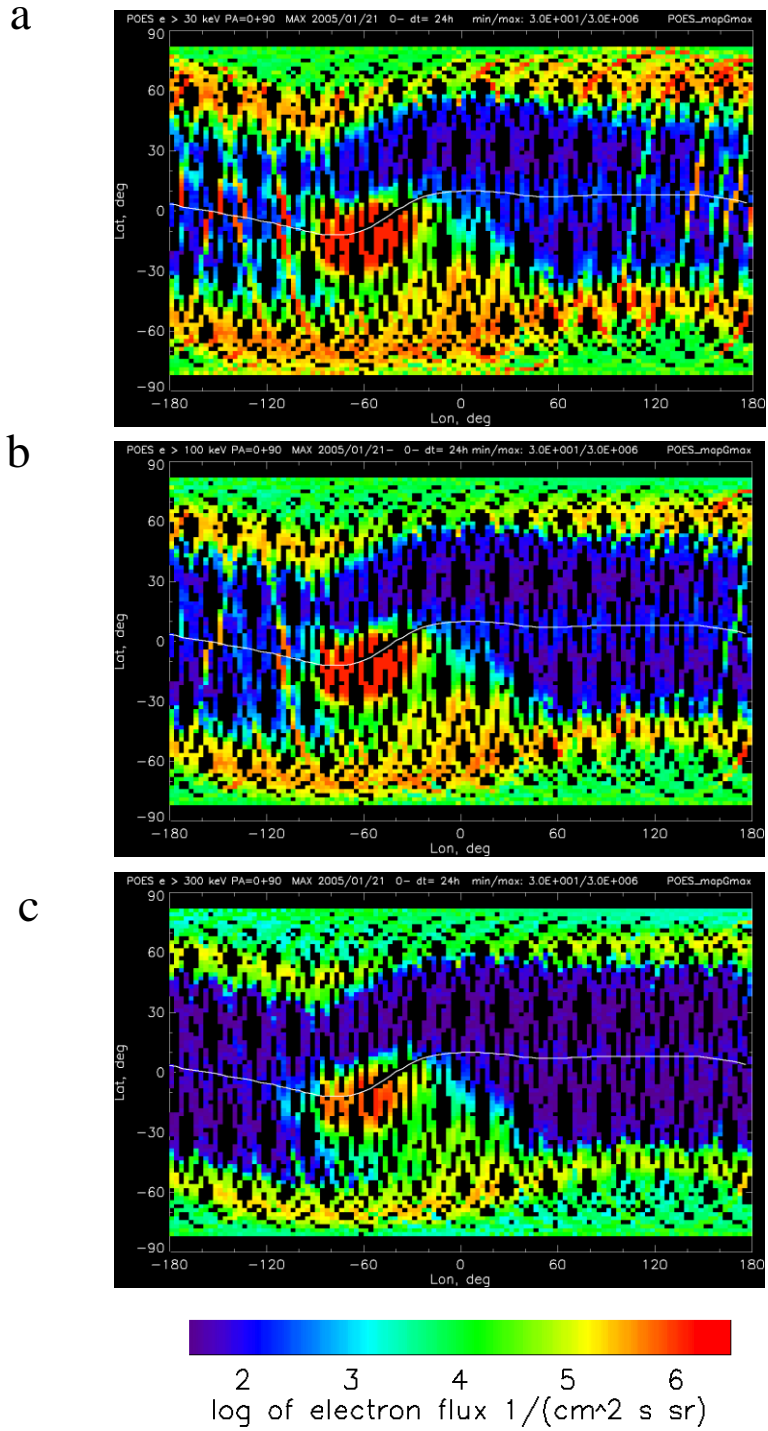

**Figure 2.** The same as Figure 1 but for extremely strong magnetospheric compression and moderate geomagnetic storm on 21 January 2005.

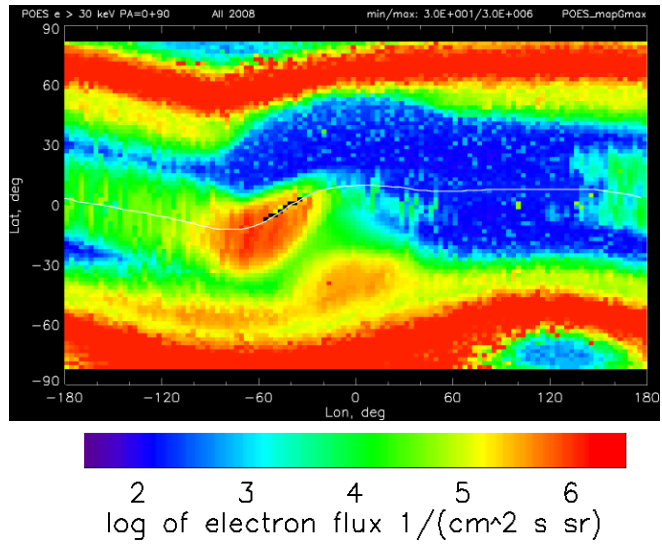

**Figure 3.** Global distribution of the electron fluxes in energy range >30 keV measured by the NOAA/POES -15, -16, -17, -18 and METOP-02 satellites during the whole year 2008 of the solar activity minimum. Low-latitude electron enhancements were observed during only 60 days in the course of the year.